\documentclass[11pt]{article}

\usepackage{fullpage}
\usepackage{setspace}
\usepackage{parskip}
\usepackage{titlesec}
\usepackage[section]{placeins}
\usepackage{xcolor}
\usepackage{breakcites}
\usepackage{lineno}
\usepackage{hyphenat}
\usepackage{rotating}
\usepackage{float}

\usepackage[margin=1in]{geometry}
\usepackage{appendix}
\usepackage{amsmath,amsfonts,bm,subfigure,enumitem}

\usepackage{tikz}
\usetikzlibrary{bayesnet}
\usetikzlibrary{arrows,shapes,snakes,automata,backgrounds,petri}

\def\vk{{\bm{k}}}

\def\vu{{\bm{u}}}
\def\vv{{\bm{v}}}

\def\vy{{\bm{y}}}
\def\vz{{\bm{z}}}

\def\mA{{\bm{A}}}
\def\mB{{\bm{B}}}

\def\mD{{\bm{D}}}

\def\mF{{\bm{F}}}
\def\mG{{\bm{G}}}
\def\mH{{\bm{H}}}

\def\mW{{\bm{W}}}

\usepackage{algpseudocode}
\usepackage{algorithm}

\PassOptionsToPackage{hyphens}{url}
\usepackage[colorlinks = true,
            linkcolor = blue,
            urlcolor  = blue,
            citecolor = blue,
            anchorcolor = blue]{hyperref}
\usepackage{etoolbox}

\usepackage{cite}
\renewenvironment{abstract}
  {{\bfseries\noindent{\abstractname}\par\nobreak}\footnotesize}
  {\bigskip}

\titlespacing{\section}{0pt}{*3}{*1}
\titlespacing{\subsection}{0pt}{*2}{*0.5}
\titlespacing{\subsubsection}{0pt}{*1.5}{0pt}

\usepackage{authblk}

\usepackage{graphicx}
\usepackage[space]{grffile}
\usepackage{latexsym}
\usepackage{textcomp}
\usepackage{longtable}
\usepackage{tabulary}
\usepackage{booktabs,array,multirow}
\usepackage{amsfonts,amsmath,amssymb}
\providecommand\citet{\cite}
\providecommand\citep{\cite}

\newif\iflatexml\latexmlfalse

\AtBeginDocument{\DeclareGraphicsExtensions{.pdf,.PDF,.png,.png,.png,.png,.tif,.TIF,.jpg,.JPG,.jpeg,.JPEG}}

\usepackage[utf8]{inputenc}
\usepackage[english]{babel}

\usepackage{float}

\iflatexml

\else

\begin{document}

\title{\bf Robust Quantitative Susceptibility Mapping via Approximate Message Passing with Parameter Estimation}

\author[1]{Shuai Huang}
\author[2]{James J. Lah}
\author[1,2]{Jason W. Allen}
\author[1]{Deqiang Qiu\thanks{This work is supported by National Institutes of Health under Grants R21AG064405, R01AG072603 and P30AG066511. Corresponding author: Deqiang Qiu (deqiang.qiu@emory.edu).}
}

\affil[1]{Department of Radiology and Imaging Sciences, Emory University, Atlanta, GA, 30322, USA}
\affil[2]{Department of Neurology, Emory University, Atlanta, GA, 30322, USA}

\vspace{-1em}

 \date{}

\begingroup
\let\center\flushleft
\let\endcenter\endflushleft
\maketitle
\endgroup

Updated final version is accepted and available in ``Magnetic Resonance in Medicine'':\\
 \urlstyle{tt}\url{https://doi.org/10.1002/mrm.29722}
\\

The code files for image reconstruction are available at:\\
\urlstyle{tt}\url{https://github.com/EmoryCN2L/QSM_AMP_PE}

\onehalfspacing

\newpage
\selectlanguage{english}
\begin{abstract}
\normalsize
{\bfseries Purpose:} For quantitative susceptibility mapping (QSM), the lack of ground-truth in clinical settings makes it challenging to determine suitable parameters for the dipole inversion. We propose a probabilistic Bayesian approach for QSM with built-in parameter estimation, and incorporate the nonlinear formulation of the dipole inversion to achieve a robust recovery of the susceptibility maps.

\par\null

{\bfseries Theory:} From a Bayesian perspective, the image wavelet coefficients are approximately sparse and modelled by the Laplace distribution. The measurement noise is modelled by a Gaussian-mixture distribution with two components, where the second component is used to model the noise outliers. Through probabilistic inference, the susceptibility map and distribution parameters can be jointly recovered using approximate message passing (AMP).

\par\null

{\bfseries Methods:} We compare our proposed AMP with built-in parameter estimation (AMP-PE) to the state-of-the-art L1-QSM, FANSI and MEDI approaches on the simulated and \emph{in vivo} datasets, and perform experiments to explore the optimal settings of AMP-PE. Reproducible code is available at \urlstyle{tt}\url{https://github.com/EmoryCN2L/QSM_AMP_PE}

\par\null

{\bfseries Results:} On the simulated Sim2Snr1 dataset, AMP-PE achieved the lowest NRMSE, DFCM and the highest SSIM, while MEDI achieved the lowest HFEN. On the \emph{in vivo} datasets, AMP-PE is robust and successfully recovers the susceptibility maps using the estimated parameters, whereas L1-QSM, FANSI and MEDI typically require additional visual fine-tuning to select or double-check working parameters.

\par\null

{\bfseries Conclusion:} AMP-PE provides automatic and adaptive parameter estimation for QSM and avoids the subjectivity from the visual fine-tuning step, making it an excellent choice for the clinical setting. 
\end{abstract}


\sloppy

\pagebreak

\section{Introduction}
\label{sec:intro}
The quantitative susceptibility mapping (QSM) technique recovers magnetic susceptibility from magnetic resonance (MR) phase images \cite{Wang:QSM:2015,Langkammer:QSM:2013,Deistung:QSM_R2Star:2013,Barbosa:QSM_R2Star:2015,Betts:QSM_R2Star:2016,Qiu1085}. It is widely used to study iron deposition in the brain \cite{Langkammer:QSM_iron:2012,Schweser:QSM:2012,Li:QSM:2011}, or pathologies such as hemorrhage \cite{Zhang:QSM_hemorrhage:2018,Sun:QSM_hemorrhage:2018} and calcification \cite{Deistung:QSM_calcification:2013,Chen:QSM_calcification:2014}. Since the values of raw phase images fall within $(-\pi,\pi]$, the phases must first be unwrapped to eliminate the discontinuity caused by the transition at $\pi$ (or $-\pi$) \cite{NonlinearMEDI:Liu:2013}. To define the brain as the region of interest (ROI), a binary brain mask can be generated using the Brain Extraction Tool \cite{Smith:BET:2002}. Magnetic field variations in the ROI are extracted from unwrapped phase images, and comprise 1) the background field induced by global geometry, air--tissue interfaces and ﬁeld inhomogeneities, and 2) the local field induced by the brain. The background field is then removed to produce the local field map \cite{Liu:PDF:2011,SCHWESER:SHARP:2011}. Recovery of the susceptibility map $\chi$ from the local field map is known as dipole inversion, an ill-posed inverse problem due to the zeros in the dipole kernel along the magic angle \cite{Wang:QSM:2015,HAACKE20151}. In this case, prior information about the susceptibility map is needed to improve the image quality. In general, image gradients are assumed to be sparse and mostly close to zero. This gives rise to the total-variation (TV) minimization approach that regularizes a data-fidelity term using the $l_1$-norm of image gradients \cite{RUDIN1992259,Strong_2003,Chambolle2004}. In QSM, the TV-minimization approach often incorporates anatomical information in the form of an edge-preserving mask obtained from the gradients of the magnitude image \cite{Liu:MEDI:2012}. Alternatively, image wavelet coefficients are also sparse in general, allowing us to select the $l_1$-norm of wavelet coefficients as the regularization term instead \cite{DBWav92,l1stable06,Yang2010ARO}. The advantage of the wavelet basis over the total-variation basis is that the wavelet transform is invertible, which offers us more freedom in developing suitable reconstruction algorithms.

Apart from the ill-posedness of the dipole inversion, phase unwrapping errors also make the problem challenging. Although the relationship between the phase and the susceptibility is linear, erroneous phase jumps in the phase image can lead to severe streaking artifacts in QSM when a linear least-squares data-fidelity term is used. To address these phase jumps, Liu et al. proposed a more robust nonlinear least-squares data-fidelity term by mapping the phases of the local field to the complex domain using the complex exponential function \cite{NonlinearMEDI:Liu:2013}. However, this non-linear data-fidelity term leads to a nonconvex problem,  making the solution dependent on initialization and susceptible to getting trapped in local minima or diverging when the input phase spans a wide dynamic range. Both the linear and nonlinear least-squares data-fidelity terms imply that the noise is modeled as additive white Gaussian noise (AWGN). However, the AWGN models are assumed in different domains: the linear data-fidelity term is in the phase domain, while the non-linear data-fidelity term is in the complex MRI signal domain. Furthermore, noise outliers can be better modeled by long-tailed distributions, such as the Gaussian-mixture distribution.

Regularization approaches, such as TV-minimization, use a parameter $\lambda$ to balance the trade-off between the data-fidelity term and the regularization term. The QSM reconstruction can have a variety of noise profiles depending on the subject’s condition and the chosen processing pipeline. When it is uncertain whether the noise profiles between the training and test sets would match, using a pre-tuned (fixed) parameter might not be ideal. As a result, researchers have turned to the L-curve method to find the parameter adaptively for each dataset \cite{Hansen:l_curve:2000,Milovic:PT_QSM:2021}. However, the L-curve method is inherently heuristic, and there is no guarantee that the selected parameter will be optimal. In practice, visual fine-tuning is typically used to find the working parameters or double-check the pre-tuned parameters for \emph{in vivo} reconstructions \cite{Milovic:L1QSM:2022}. However, the parameters chosen through visual fine-tuning are subjective and depend on the practitioner.

In this paper, we propose a probabilistic Bayesian approach to jointly recover the susceptibility map and parameters. We use the Laplace distribution to encode the sparse prior on the wavelet coefficients of susceptibility map, and a customized Gaussian-mixture distribution to model the noise distribution. We compute the maximum-a-posteriori (MAP) estimations of the wavelet coefficients and distribution parameters using approximate message passing (AMP)\cite{Rangan:GAMP:2011,PE_GAMP17}. To handle the phase unwrapping errors, we adopt the nonlinear measurement model where the complex exponential functions of phases are used as measurements. We then extend the standard linear AMP so that it could be used to solve the nonlinear dipole inversion. In addition, we propose a morphology mask for the image wavelet coefficients to incorporate anatomical structural information into the reconstruction. Experiments show that the proposed AMP with built-in parameter estimation (AMP-PE) is robust and successfully recovers susceptibility maps on both simulated and \emph{in vivo} datasets.

\section{Theory}

Let $\mB_l$ denote the produced local field after phase unwrapping and background field removal, and $\boldsymbol\phi_e$ denote the corresponding phase at an echo time $t_e$. We have
\begin{align}
\label{eq:linear_measurement}
\begin{split}
    \boldsymbol\phi_e&=2\pi\gamma\cdot t_e\cdot \mB_l+\widetilde{\vu}+\boldsymbol\phi_0\\ 
    &=2\pi\gamma\cdot t_e\cdot B_0\cdot\mF^*\mD\mF\boldsymbol\chi+\widetilde{\vu}+\boldsymbol\phi_0\\
    &=\mA_e\boldsymbol\chi+\widetilde{\vu}+\boldsymbol\phi_0\,,
\end{split}
\end{align}
where $\gamma$ is the gyromagnetic ratio, $\boldsymbol\phi_0$ is the initial phase offset that depends on the coil-sensitivity, $B_0$ is the main magnetic field, $\mF$ is the Fourier transform matrix, $\mD$ is the dipole kernel in the frequency domain, $\boldsymbol\chi$ is the magnetic susceptibility, $\widetilde{\vu}$ is the noise, and $\mA_e$ is the (combined) resulting operator applied on $\boldsymbol\chi$. The initial phase $\boldsymbol\phi_0$ can be estimated from the multi-echo phase images by solving a nonlinear least-squares fitting problem \cite{NonlinearMEDI:Liu:2013}. In particular, the ill-posed dipole kernel $\mD$ is given by
\begin{align}
    \mD(\vk)=\left\{\begin{array}{c}
    \frac{1}{3}-\frac{k_z^2}{\|\vk\|_2^2}\\
    0
    \end{array}
    \quad
    \begin{array}{l}
    \textnormal{if }\vk\neq\boldsymbol 0\\
    \textnormal{if }\vk=\boldsymbol 0\,,
    \end{array}
    \right.
\end{align}
where $\vk=\left[k_x\ k_y\ k_z\right]^T$ is the spatial frequency component. 

To achieve a robust recovery of the susceptibility $\boldsymbol\chi$, we adopt the following nonlinear forward model proposed in \cite{NonlinearMEDI:Liu:2013}:
\begin{align}
\label{eq:exp_measurement_model}
\mW_e\cdot \exp(i\boldsymbol\phi_e) = \mW_e\cdot\exp(i\mA_e\boldsymbol\chi)+\vu\,,
\end{align}
where $i$ is the imaginary unit, $\mW_e$ is a diagonal weighting matrix, $\vu$ is the noise. The complex exponential function $\exp(i\boldsymbol\phi_e)$ of the phase image $\boldsymbol\phi_e$ could effectively reduce the discontinuity caused by erroneous phase jumps \cite{NonlinearMEDI:Liu:2013}. However, the nonlinear forward model makes the problem nonconvex, and the solution could get trapped in some local minima or diverge if it's not properly initialized. There are various methods for designing suitable weighting matrices, but in this case, we can simply use the magnitude image as $\mW_e$. Although the AWGN model is loosened when the weighting matrix $\mW_e$ is used, experiments demonstrate that the proposed AMP-PE approach can still adjust the estimated noise variance accordingly to fit the loosened AWGN model for a successful recovery.

\subsection{Bayesian Formulation}
\begin{figure}[tbp]
\begin{center}
\subfigure[Laplace distribution]{
\label{fig:laplace_signal}
\includegraphics[width=.45\textwidth]{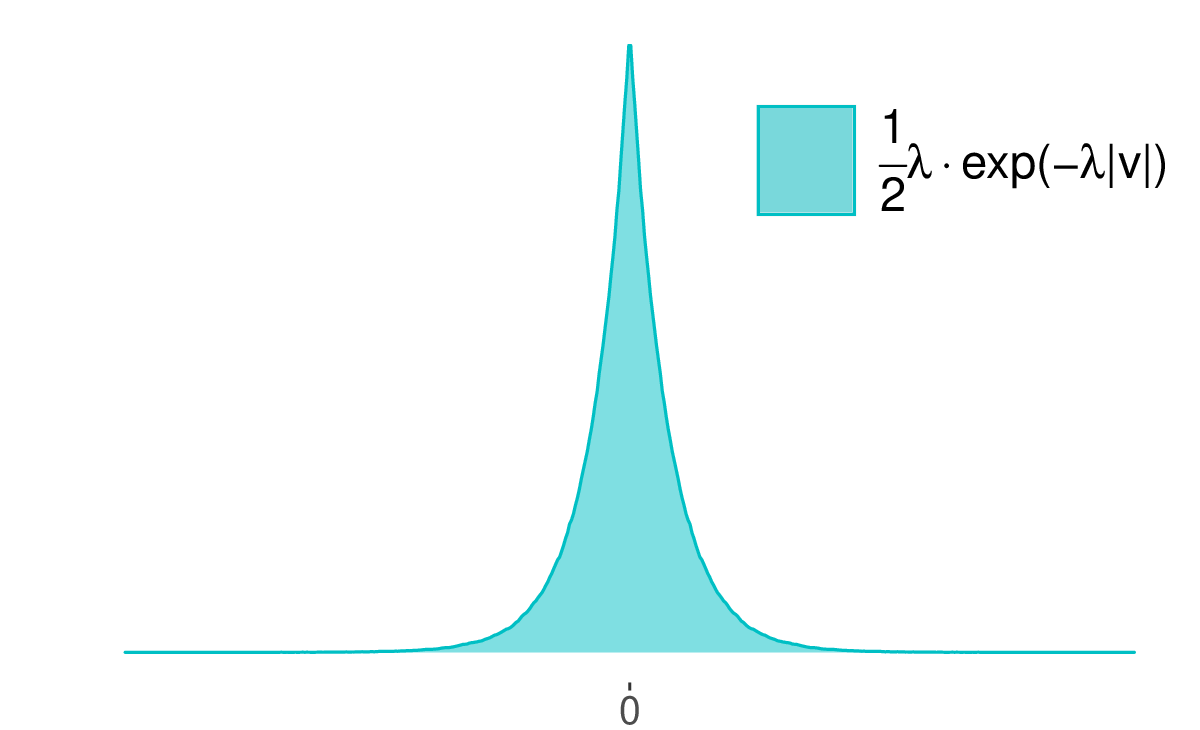}}
\subfigure[Gaussian-mixture distribution]{
\label{fig:gm_noise}
\includegraphics[width=.45\textwidth]{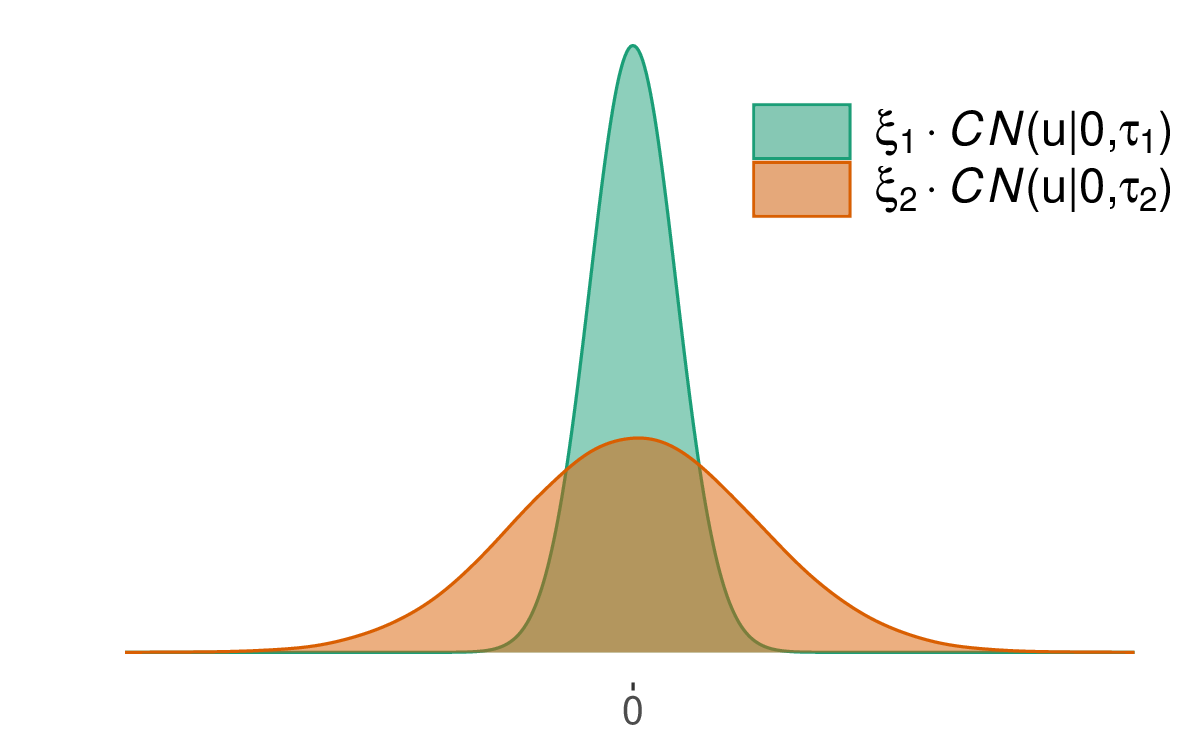}}
\end{center}
\caption{The prior distributions in the probabilistic Bayesian formulation: (a) the image wavelet coefficients are sparse and modeled by the Laplace distribution; (b) the noise is modeled by the Gaussian-mixture distribution with two components, the second Gaussian component is used to model the noise outliers.}
\label{fig:laplace_gm_priors}
\end{figure}

The sparse prior on the wavelet coefficients $\vv$ of the susceptibility $\boldsymbol\chi$ is used to improve the image quality:
\begin{align}
    \vv=\mH\boldsymbol\chi\,,
\end{align}
where $\mH$ is the invertible wavelet transform matrix. As shown in Fig. \ref{fig:laplace_signal}, we use the sparsity-promoting Laplace distribution to model the distribution of wavelet coefficients. The coefficients in $\vv$ are assumed to be independent and identically distributed (i.i.d.):
\begin{align}
\label{eq:laplace_prior}
    p(v|\lambda) = \frac{1}{2}\lambda\cdot\exp(-\lambda|v|)\,,
\end{align}
where $\lambda>0$ is the unknown distribution parameter. 

As shown in Fig. \ref{fig:gm_noise}, we propose the following Gaussian-mixture distribution with two components to model the noise distribution
\begin{align}
\label{eq:gm_noise_prior}
    p(u|\xi_s,\tau_s)=\sum_{s=1}^2\xi_s \cdot\mathcal{CN}(u|0,\tau_s),
\end{align}
where $\xi_s$ is the $s$-th mixture weight, $\tau_s$ is the $s$-th variance, the mixture means are zeros, and $\mathcal{CN}(\cdot)$ is the complex Gaussian density function. The number of Gaussian mixtures needs to be chosen carefully to avoid overfitting during reconstruction. From the experiments later in Section \ref{sec:results}, we observe that using two Gaussian mixtures produces the best performance. The second Gaussian component $\mathcal{CN}(u|0,\tau_2)$ is used to model the noise outliers that give rise to a long-tailed distribution. The variance $\tau_2$ of the second Gaussian component should be large enough to cover the domain of $u$. In practice, we can initialize the second variance $\tau_2$ with a larger value than the first variance $\tau_1$.

Under the Bayesian formulation, we can recover $\boldsymbol\chi$ by computing its MAP estimation
\begin{align}
    \widehat{\boldsymbol\chi} = \arg\max_{\boldsymbol\chi}\ p(\boldsymbol\chi|\vy)\,,
\end{align}
where $\vy$ contains the measurements, and $p(\boldsymbol\chi|\vy)$ is the posterior distribution of $\boldsymbol\chi$ that can be computed via approximate message passing (AMP) \cite{Rangan:GAMP:2011}.

\subsection{Approximate Message Passing with Parameter Estimation}
\label{subsec:amp_pe}
AMP was originally designed for the linear measurement system \cite{Rangan:GAMP:2011}. In order to apply it for the nonlinear dipole inversion, we need to linearize it iteratively using the zero-th and first orders of Taylor series \cite{NonlinearMEDI:Liu:2013}. Letting $\boldsymbol\chi^{(r)}$ denote the susceptibility in the $r$-th iteration, we have the following linear approximation in the $(r+1)$ iteration:
\begin{align}
\label{eq:exp_measurement_model_linearized}
    \mW_e\cdot\exp(i\boldsymbol\phi_e) = i\mW_e\cdot\exp\left(i\mA_e\boldsymbol\chi^{(r)}\right)\cdot\mA_e\boldsymbol\chi - g\left(\boldsymbol\chi^{(r)}\right) + \vu\,,
\end{align}
where $g\left(\boldsymbol\chi^{(r)}\right)=\mW_e\cdot\exp\left(i\boldsymbol\mA_e\boldsymbol\chi^{(r)}\right)\cdot\left(i\mA_e\boldsymbol\chi^{(r)}-\boldsymbol 1\right)$ is a relative constant that depends on $\boldsymbol\chi^{(r)}$. Rewriting the above \eqref{eq:exp_measurement_model_linearized} in the form of a linear measurement model, we have
\begin{align}
\label{eq:exp_measurement_model_linearized_rewrite}
\mW_e\cdot\exp(i\boldsymbol\phi_e) + g\left(\boldsymbol\chi^{(r)}\right) = i\mW_e\cdot\exp\left(i\mA_e\boldsymbol\chi^{(r)}\right)\cdot\mA_e\boldsymbol\chi  + \vu\,,
\end{align}
where $\mW_e\cdot\exp(i\boldsymbol\phi_e) + g\left(\boldsymbol\chi^{(r)}\right)$ become the linear measurements, and $i\mW_e\cdot\exp\left(i\mA_e\boldsymbol\chi^{(r)}\right)\cdot\mA_e$ becomes the corresponding measurement matrix. We can then recover $\boldsymbol\chi^{(r+1)}$ using AMP based on the linear approximation in \eqref{eq:exp_measurement_model_linearized_rewrite}.

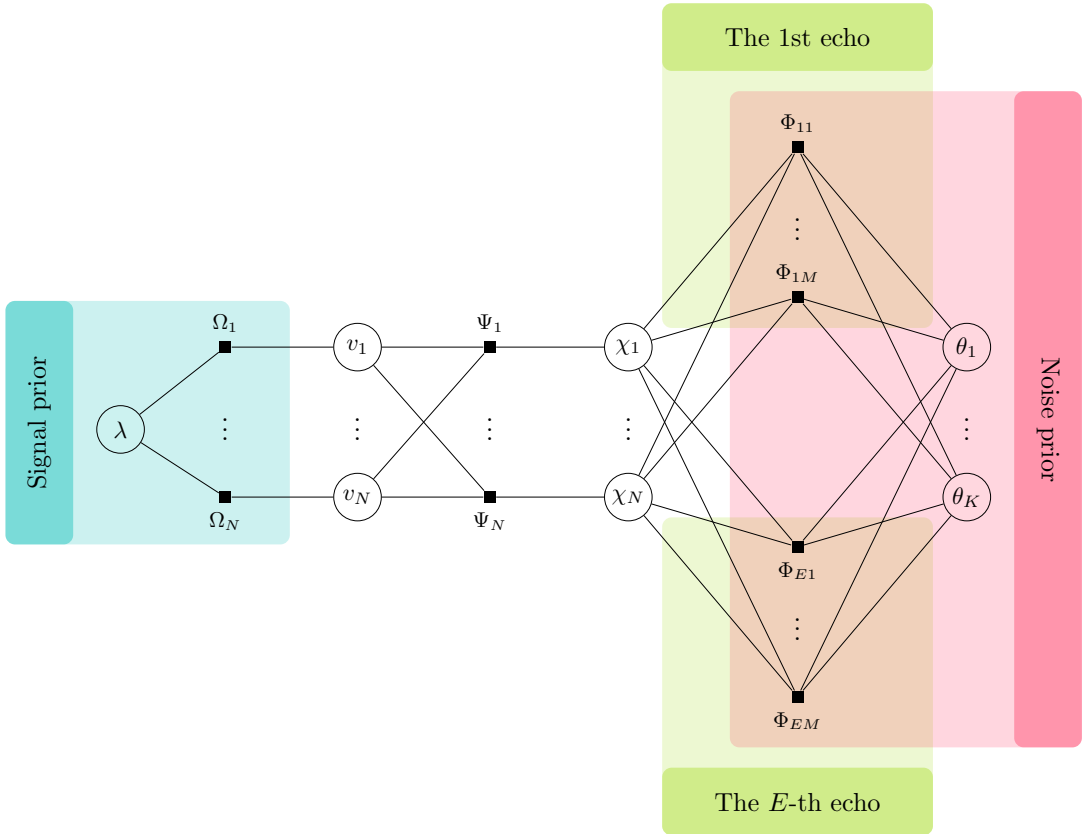
\begin{figure}[tbp]
\begin{center}
\begin{tabular}{c}
\scalebox{0.9}{
%
%
%
%
\definecolor{00BCB4}{RGB}{0,188,180}
\definecolor{C4E86B}{RGB}{196,232,107}
\definecolor{FF3561}{RGB}{255,53,97}

\begin{tikzpicture}

  
  \fill[00BCB4, opacity=0.2, rounded corners] (-1.7,-1.7) rectangle (2.5,1.9);
  \fill[C4E86B, opacity=0.3, rounded corners] (8,1.5) rectangle (12,6.3);
  \fill[C4E86B, opacity=0.3, rounded corners] (8,-6) rectangle (12,-1.3);
  \fill[FF3561, opacity=0.2, rounded corners] (9,-4.7) rectangle (14.2,5);

  \fill[00BCB4, opacity=0.4, rounded corners] (-1.7,-1.7) rectangle (-0.7,1.9);
  \node[text width=3cm, rotate = 90, align=center] at (-1.2,0.1) {Signal prior};
  \fill[C4E86B, opacity=0.7, rounded corners] (8,5.3) rectangle (12,6.3);
  \node[text width=4.5cm, align=center] at (10,5.8) {The 1st echo};
  \fill[C4E86B, opacity=0.7, rounded corners] (8,-6) rectangle (12,-5);
  \node[text width=4.5cm, align=center] at (10,-5.5) {The $E$-th echo};
  \fill[FF3561, opacity=0.4, rounded corners] (13.2,-4.7) rectangle (14.2,5);
  \node[text width=3cm, rotate = -90, align=center] at (13.7,0.15) {Noise prior};

  \node[latent,fill=00BCB4!20] (lambda) {$\lambda$};
  
  \node[latent, above=0.5 of lambda, xshift=3.5cm] (v_1) {$v_1$};
  \node[latent, below=1.5 of v_1] (v_N) {$v_N$};

  \path (v_1) -- node[auto=false]{\vdots} (v_N);
  
  \factor[left=0 of v_1, xshift=-1.5cm] {Omega_1-f} {above:$\Omega_1$} {v_1, lambda}{};
  \factor[left=0 of v_N, xshift=-1.5cm] {Omega_N-f} {below:$\Omega_N$} {v_N, lambda}{};
  \path(Omega_1-f) -- node[auto=false]{\vdots} (Omega_N-f);
  
  \node[latent, above=0.5 of lambda, xshift=7.5cm] (x_1) {$\chi_1$};
  \node[latent, below=1.5 of x_1] (x_N) {$\chi_N$};
  
  \path (x_1) -- node[auto=false]{\vdots} (x_N);
  
  \factor[right=0 of v_1, xshift=1.5cm] {Psi_1-f} {above:$\Psi_1$} {x_1, v_1, v_N} {};
  \factor[right=0 of v_N, xshift=1.5cm] {Psi_N-f} {below:$\Psi_N$} {x_N, v_1, v_N} {};
  
  \path(Psi_1-f) -- node[auto=false]{\vdots} (Psi_N-f);

  \node[latent,fill=FF3561!20, above = 0.5 of lambda, xshift=12.5cm] (theta_1) {$\theta_1$};
  \node[latent,fill=FF3561!20, below = 1.5 of theta_1] (theta_K) {$\theta_K$};
  \path(theta_1) -- node[auto=false]{\vdots} (theta_K);

  \factor[right=0 of x_1, above = 2.5 of x_1, xshift=2.5cm] {Phi_11-f} {above:$\Phi_{11}$} {theta_1, theta_K, x_1, x_N} {};
  \factor[right=0 of x_N, above = 2.5 of x_N, xshift=2.5cm] {Phi_1M-f} {above:$\Phi_{1M}$} {theta_1, theta_K, x_1, x_N} {};
  
  \path(Phi_11-f) -- node[auto=false]{\vdots} (Phi_1M-f);

  \factor[right=0 of x_1, below = 2.5 of x_1, xshift=2.5cm] {Phi_E1-f} {below:$\Phi_{E1}$} {theta_1, theta_K, x_1, x_N} {};
  \factor[right=0 of x_N, below = 2.5 of x_N, xshift=2.5cm] {Phi_EM-f} {below:$\Phi_{EM}$} {theta_1, theta_K, x_1, x_N} {};
  
  \path(Phi_E1-f) -- node[auto=false]{\vdots} (Phi_EM-f);
  
\end{tikzpicture}

}
\end{tabular}
\end{center}
\caption{The factor graph of the QSM task: ``$\bigcirc$'' represents the variable node, and ``$\blacksquare$'' represents the factor node. In particular, $\lambda$ is the signal prior parameter, $\boldsymbol\theta=\{\xi_1,\xi_2,\tau_1,\tau_2\}$ contains the noise prior parameters.}
\label{fig:factor_graph_qsm}
\end{figure}

As shown by the factor graph in Fig. \ref{fig:factor_graph_qsm}, the distribution parameters $\{\lambda,\boldsymbol\theta\}$ are treated as random variables and jointly recovered with the signals of interest $\{\vv, \boldsymbol\chi\}$ \cite{PE_GAMP17}. The variable nodes are represented by ``$\bigcirc$'' and contain random variables. The factor nodes are represented by ``$\blacksquare$'' and encode the probability distributions of random variables. The messages about the variable distributions are passed and discussed among the factor nodes until a consensus is reached. 

For example, we use the following notations to denote the messages between the $n$-th variable node $\chi_n$ and the $m$-th factor node $\Phi_{em}$ in the $e$-th echo:
\begin{itemize}
    \item $\Delta_{\chi_n\rightarrow\Phi_{em}}$ denotes the message from $\chi_n$ to $\Phi_{em}$,
    \item $\Delta_{\Phi_{em}\rightarrow \chi_n}$ denotes the message from $\Phi_{em}$ to $\chi_n$,
\end{itemize}
where $n\in\{1,\cdots,N\}$, $e\in\{1,\cdots,E\}$ and $m\in\{1,\cdots,M\}$. Both $\Delta_{\chi_n\rightarrow\Phi_{em}}$ and $\Delta_{\Phi_{em}\rightarrow \chi_n}$ are functions of the variable $\chi_n$, and are expressed in the ``$\log$'' domain in this paper. A derivation of the AMP algorithm is beyond the scope of this paper, and algorithmic details can be found in \cite{Rangan:GAMP:2011,PE_GAMP17}. A detailed introduction to AMP is included in Section S-I of the Supporting Information. 

Once the message passing converges, we can calculate the distributions of the variables using the messages. For example, the posterior distribution $p(\chi_n|\vy)$ is proportional to the exponential function of the summation of all the messages passed to $\chi_n$ \cite{Wainwright:Graph:2008,Minka:2001}:
\begin{align}
p(\chi_n|\vy)\propto\exp\left(\Delta_{\Psi_n\rightarrow\chi_n}+\sum_{em}\Delta_{\Phi_{em}\rightarrow\chi_n}\right)\,.
\end{align}

Currently, the convergence of the AMP algorithm has only been established for random Gaussian measurement matrices \cite{Rangan:GAMP:2011}. Extending the convergence analysis to general measurement matrices is still an open question. In practice, damping and mean-removal operations are used to ensure the convergence of AMP for non-Gaussian matrices \cite{Rangan:DampingCvg:2014,Vila:DampingMR:2015}. For the ill-posed measurement operator in \eqref{eq:exp_measurement_model_linearized}, which contains the dipole kernel, we perform the damping operation on $\boldsymbol\chi$ to stabilize AMP. Let $\boldsymbol\chi_d^{(t)}$ denote the damped susceptibility in the $t$-th iteration, and $\boldsymbol\chi^{(t+1)}$ denote the undamped susceptibility in the $(t+1)$-th iteration. The MAP estimation of the $n$-th entry $\chi_n$ in $\boldsymbol\chi$ is
\begin{align}
    \chi_n^{(t+1)} = \arg\max_{\chi_n}\ p(\chi_n|\vy) = \arg\max_{\chi_n}\ \Delta^{(t+1)}_{\Psi_n\rightarrow\chi_n}+\sum_{em}\Delta^{(t+1)}_{\Phi_{em}\rightarrow\chi_n}\,.
\end{align}
We can compute the damped solution $\boldsymbol\chi_d^{(t+1)}$ as follows
\begin{align}
\label{eq:damping_chi}
    \boldsymbol\chi_d^{(t+1)} = \boldsymbol\chi_d^{(t)}+\alpha\cdot\left(\boldsymbol\chi^{(t+1)}-\boldsymbol\chi_d^{(t)}\right)\,,
\end{align}
where $\alpha\in(0,1]$ is the damping rate on the susceptibility $\boldsymbol\chi$. For the QSM task, we need to choose a small damping rate $\alpha$ to be around $0.01$.

As discussed in \cite{PE_GAMP17}, we can estimate the parameter $\lambda$ of the sparse signal prior by maximizing its posterior:
\begin{align}
\label{eq:map_lambda}
    \widehat{\lambda} = \arg\max_{\lambda}\ p(\lambda|\vy) = \arg\max_{\lambda}\ \sum_n\Delta_{\Omega_n\rightarrow\lambda} \,,
\end{align}
where the posterior distribution $p(\lambda|\vy)$ can be computed via AMP.

The estimation of the noise prior parameters requires additional work. Due to the ill-posedness of the measurement operator, the MAP method tends to over-estimate the weight $\xi_2$ of the second Gaussian component, which is reserved for the noise outliers. To overcome this issue, we propose a two-step procedure to estimate the mixture weights $\xi_1,\xi_2$:
\begin{enumerate}[label=\arabic*)]
    \item Perform a preliminary reconstruction with only one Gaussian component $\mathcal{CN}(u|0,\tau_0)$ to model the noise.
    \begin{align}
        \widehat{\tau}_0 = \arg\max_{\tau_0}\ p(\tau_0|\vy) = \arg\max_{\tau_0}\ \sum_{em}\Delta_{\Phi_{em}\rightarrow\tau_0}\,.
    \end{align}
    \item Estimate the weights $\{\xi_1,\xi_2\}$ of the two-component Gaussian mixture based on the preliminary reconstruction $\boldsymbol\chi_0$. We first calculate the residual error $\boldsymbol\epsilon$ using $\boldsymbol\chi_0$:
    \begin{align}
        \boldsymbol\epsilon_e = \mW_e\exp(i\boldsymbol\phi_e)-\mW_e\exp(i\mA_e\boldsymbol\chi_0)\,.
    \end{align}
    We next use $\boldsymbol\epsilon_e$ to approximate the noise $\vu$, and treat the residue entries that fall outside $[-3\tau_0,3\tau_0]$ as outliers. The weights $\xi_1$ and $\xi_2$ can then be estimated as
    \begin{align}
        \widehat{\xi}_1&=\frac{1}{EM}\sum_{em}\boldsymbol 1(|\epsilon_{em}|\leq 3\tau_0),\\
        \widehat{\xi}_2&=1-\widehat{\xi}_1\,,
    \end{align}
    where $\boldsymbol 1(\cdot)$ is the indicator function.
\end{enumerate}

With the estimated weights $\widehat{\xi}_1,\widehat{\xi}_2$ fixed, we can estimate the Gaussian mixture variances $\tau_1,\tau_2$ by maximizing their posteriors:
\begin{align}
\label{eq:map_tau_s}
    \widehat{\tau}_s=\arg\max_{\tau_s}\ p(\tau_s|\vy)=\arg\max_{\tau_s}\ \sum_{em}\Delta_{\Phi_{em}\rightarrow\tau_s},\quad s=1,2.
\end{align}

When computing the MAP estimations of the parameters $\lambda,\tau_1,\tau_2$, we also need to use damping operations to overcome the ill-posedness of the dipole kernel and stabilize AMP. In the $(t+1)$-th iteration, we have
\begin{align}
\label{eq:damping_lambda}
    \widehat{\lambda}_d^{(t+1)}&=\widehat{\lambda}_d^{(t)}+\beta\cdot\left(\widehat{\lambda}^{(t+1)}-\widehat{\lambda}_d^{(t)}\right),\\
\label{eq:damping_tau}
    {{}\widehat{\tau}_s}_d^{(t+1)}&={{}\widehat{\tau}_s}_d^{(t)}+\beta\cdot\left(\widehat{\tau}_s^{(t+1)}-{{}\widehat{\tau}_s}_d^{(t)}\right),
\end{align}
where $\widehat{\lambda}_d^{(t)},{{}\widehat{\tau}_s}_d^{(t)}$ are the damped parameters in the $(t)$-th iteration, $\widehat{\lambda}^{(t+1)}$ and ${\widehat{\tau}_s}^{(t+1)}$ are the undamped MAP estimations computed via \eqref{eq:map_lambda} and \eqref{eq:map_tau_s} in the $(t+1)$-th iteration, $\beta\in(0,1]$ is the damping rate on the estimated parameters. For the QSM task, we can choose $\beta$ to be around $0.1$.

\subsection{Morphology Mask for the Wavelet Coefficients}
\label{subsec:mask_wavelet_coeff}
\begin{figure}[tbp]
\begin{center}
\includegraphics[width=.7\textwidth]{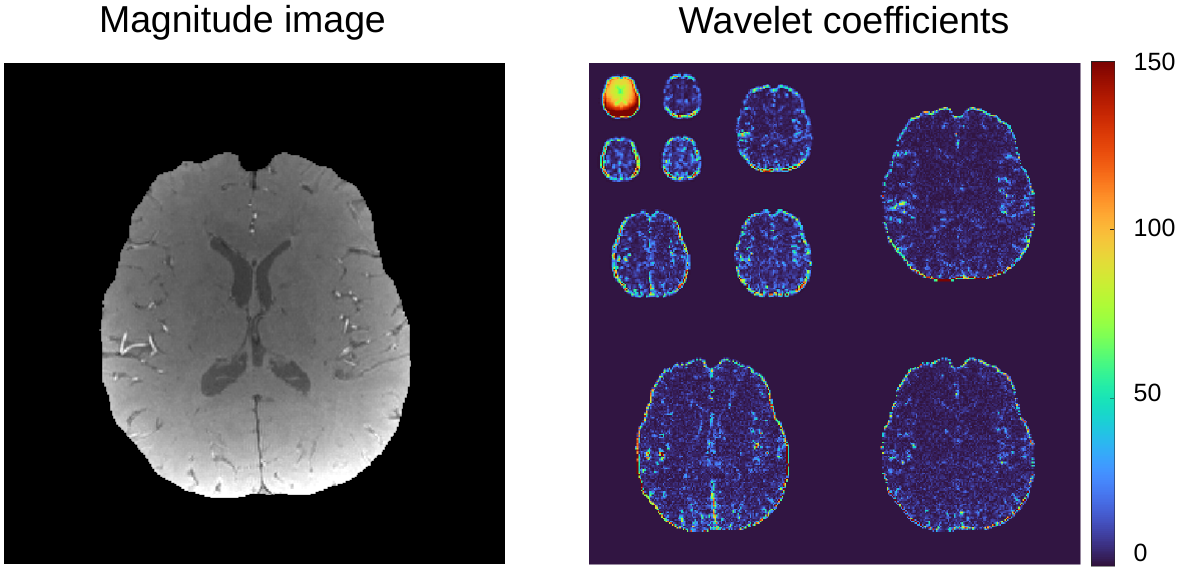}
\end{center}
\caption{The magnitude image and its wavelet coefficients obtained using the db2 wavelet basis with 3 levels of decomposition.}
\label{fig:wavelet_coefficients}
\end{figure}

To incorporate anatomical information into the reconstruction process, we propose a new morphology mask $\mathcal{M}_v$ that can be applied on the wavelet coefficients $\vv$ of the susceptibility map $\boldsymbol\chi$. The purpose of using a morphology mask is to preserve the high-frequency information that corresponds to the edges of anatomical structures. The wavelet transform applies a series of low-pass and high-pass filters on the image, generating the wavelet coefficients that provide a natural way of identifying edge information. As shown in Fig. \ref{fig:wavelet_coefficients}, the wavelet coefficients $\vz$ of the magnitude image reveal the anatomical structures in a hierarchical manner, and those significant coefficients correspond to the structural edges. Since the anatomical structures are consistent between the magnitude image and the susceptibility map, the indices of significant wavelet coefficients should also be consistent between $\vz$ of the magnitude image and $\vv$ of the susceptibility map. We can set a threshold $\mu$ on $\vz$ to generate the morphology mask $\mathcal{M}_v$, which we can then apply on $\vv$ to avoid penalizing significant coefficients during reconstruction. 

Let $\mathcal{S}$ denote the set of indices of the wavelet coefficients in $\vz\in\mathbb{R}^N$ whose absolute values are larger than the magnitude threshold $\mu$:
\begin{align}
\mathcal{S}=\{i\ |\ \left|z(i)\right|>\mu,\ \textnormal{where}\ 1\leq i\leq N\}\,.
\end{align}
The value of $\mu$ is selected with respect to the $l_1$-norm $\|\vz\|_1$ such that 
\begin{align}
\label{eq:wavelet_mask_threshold_c}
\frac{\sum_{i\in\mathcal{S}} \left|z(i)\right|}{\|\vz\|_1}=c\,,
\end{align}
where $0<c<1$ is the percentage threshold, which is defined by the percentage of top wavelet coefficients with respect to $\|\vz\|_1$. A larger $c$ leads to more high-frequency information in the recovered susceptibility map, while a smaller $c$ leads to more low-frequency information. In practice, we can set $c$ to be around 0.85 for \emph{in vivo} reconstructions. 

The binary $0-1$ morphology mask $\mathcal{M}_v$ for the wavelet coefficients $\vv$ is then
\begin{align}
\mathcal{M}_v(i)=\left\{
\begin{array}{l}
1\\
0
\end{array}
\quad
\begin{array}{l}
\textnormal{if }i\in\mathcal{S}\\
\textnormal{otherwise}.
\end{array}
\right.
\end{align}

When the wavelet coefficients $\vv$ of the susceptibility map are modeled by the Laplace distribution as in \eqref{eq:laplace_prior}, the MAP estimation of $\vv$ under the AMP formulation is obtained through the soft-thresholding operator $\mathcal{T}_{\kappa}(\cdot)$ \cite{Bellili:AMP:2019}. 
\begin{align}
\mathcal{T}_{\kappa}(v) = \left\{
\begin{array}{l}
(|v|-\kappa)\cdot \textnormal{sign}(v)\\
0
\end{array}
\quad
\begin{array}{l}
\textnormal{if }|v|>\kappa\\
\textnormal{if }|v|\leq \kappa,
\end{array}
\right.
\end{align}
where $\kappa$ is the soft threshold specified by the AMP algorithm. The soft-thresholding operator $\mathcal{T}_{\kappa}(\cdot)$ essentially penalizes the magnitude of individual coefficient. 

By utilizing the morphology mask $\mathcal{M}_v$, we ensure that coefficients belonging to the set $\mathcal{S}$ are not penalized, which helps preserving high-frequency edge information. Let $\vv^{(t)}$ denote the solution from the $(t)$-th iteration. In the $(t+1)$-th iteration, we then have
\begin{align}
v^{(t+1)}(i) = \left\{
\begin{array}{l}
v^{(t)}(i)\\
\mathcal{T}_{\kappa}\left(v^{(t)}(i)\right)
\end{array}
\quad
\begin{array}{l}
\textnormal{if }i\in\mathcal{S}\\
\textnormal{otherwise}.
\end{array}\quad 
\forall\ 1\leq i\leq N\,.
\right.
\end{align}

\section{Methods}
We compare the proposed AMP-PE approach with the state-of-the-art L1-QSM \cite{Milovic:L1QSM:2022}, FANSI \cite{Milovic:FANSI:2018,Milovic:PT_QSM:2021} and MEDI \cite{NonlinearMEDI:Liu:2013} approaches on the 2019 QSM challenge 2.0 dataset and the \emph{in vivo} 3D brain datasets. 

Specifically, the L1-QSM, FANSI and MEDI approaches adopt the total-variation regularization and solve the following nonlinear dipole inversion problems:
\begin{align}
    \label{eq:l1_qsm_obj}
    \textnormal{L1-QSM:}&\quad\min_{\boldsymbol\chi}\ \left\|\mW_e\left(\exp\left(i\mA_e\boldsymbol\chi\right)-\exp\left(i\boldsymbol\phi_e\right)\right)\right\|_1+\eta\cdot\|\mG\boldsymbol\chi\|_1,\\
    \label{eq:fansi}
    \textnormal{FANSI:}&\quad\min_{\boldsymbol\chi}\ \left\|\mW_e\left(\exp\left(i\mA_e\boldsymbol\chi\right)-\exp\left(i\boldsymbol\phi_e\right)\right)\right\|_2^2+\zeta\cdot\|\mG\boldsymbol\chi\|_1,\\
    \label{eq:medi_obj}
    \textnormal{MEDI:}&\quad\min_{\boldsymbol\chi}\ \rho\cdot\left\|\widehat{\mW}_e\left(\exp\left(i\mA_e\boldsymbol\chi\right)-\exp\left(i\boldsymbol\phi_e\right)\right)\right\|_2^2+\|\mathcal{M}_g\mG\boldsymbol\chi\|_1,
\end{align}
where $\mG(\cdot)$ is the gradient operator that computes image gradients, $\mathcal{M}_g$ is a 0-1 morphology mask and only selects the gradients that correspond to non-edges; $\eta$, $\zeta$ and $\rho$ are the regularization parameters. The L1-QSM and FANSI approaches use the magnitude image as the weighting matrix $\mW_e$, while the MEDI approach constructs $\widehat{\mW}_e$ based on the reliability of the measurements and iteratively updates it according to the residual. 
In order to showcase the robustness of using complex exponential measurements $\mW_e\exp(i\boldsymbol\phi_e)$, we also recover QSM using the conventional linear measurements $\mW_e\boldsymbol\phi_e$. The results from the linear recovery approach are given in Section S-V of the Supporting Information.


In the clinical setting, we could not perform parameter-tuning for QSM due to the lack of ground-truth across different processing pipelines, acquisition protocols and scanners. For the FANSI approach, we followed the guidelines in \cite{Milovic:PT_QSM:2021} and used the heuristic L-curve method to determine the parameter $\zeta$ from the set $\{10^{(-1.5-i*0.1)}\ |\ i=1,\cdots,25\}$, as suggested by the script provided in the FANSI toolbox. The maximum number of iterations was set to 300, with a convergence rate of $1e^{-3}$. Additionally, we set the Lagrangian weight in the total-variation subproblem to $\mu_1=100\zeta$, and the weight in the data-fidelity term subproblem to $\mu_2=1$. The curvature data of the L-curve was smoothed using a median filter, and the inflection point where the curvature changes sign was selected as the parameter. If the L-curve method fails to produce a suitable parameter, we perform visual fine-tuning. For the L1-QSM approach, we followed the same guidelines as the FANSI approach. The optimal parameter $\eta$ is generally larger than that of FANSI, and is thus selected from the set $\{10^{(-0.5-i*0.1)}\ |\ i=1,\cdots,20\}$. For the MEDI approach, the L-curve method cannot be used in this case since the weighting matrix $\mW_e$ in \eqref{eq:medi_obj} is iteratively updated with respect to the residue, which introduces additional variability to the data-fidelity term that alters the shape of the L-curve. Therefore, we used the tuned parameter $\rho=1000$, as suggested by the MEDI-toolbox \cite{NonlinearMEDI:Liu:2013}. 

Unlike other approaches that use the total-variation regularization, AMP-PE uses a sparse prior on the image wavelet coefficients to improve image quality. We use the Daubechies wavelet family to obtain the sparse representation of the susceptibility map \cite{DBWav92}. The orthogonal ``db1--db10'' wavelet bases are typically used, with the complexity of a wavelet basis increasing with its order. For image recovery, we use a wavelet transform with three levels of decomposition. Specifically, the db1 wavelet is suitable for capturing low-frequency information in the image, while higher-order wavelet bases are better suited for capturing high-frequency information. Although we aim to retain the structural details of the susceptibility map encoded in the high-frequency bands, streaking artifacts also contain predominantly high-frequency information. Therefore, we need to find a suitable basis that balances the trade-off between low and high frequencies. As discussed in Section \ref{subsec:mask_wavelet_coeff}, we apply the proposed morphology mask $\mathcal{M}_v$ on the wavelet coefficients to incorporate anatomical information into the reconstruction. The mask $\mathcal{M}_v$ is determined by the percentage threshold $c$ in \eqref{eq:wavelet_mask_threshold_c}. The larger the threshold $c$ is, the more high-frequency information will be retained in the susceptibility map.
\begin{itemize}
\item For the simulated Sim2Snr1 dataset, much of the ground-truth susceptibility map is piecewise constant and contains more low-frequency information. Thus, the db1 wavelet is a better choice in this case, and the percentage threshold $c$ is set to $0.75$.
\item For the \emph{in vivo} datasets, the susceptibility map contains a lot of fine structural details that cannot be fully captured by the db1 wavelet, and the more complex db2 wavelet basis can be chosen instead. Higher-order wavelet bases such as db3 could capture more details from the streaking artifacts, and are generally not recommended for the QSM task. The percentage threshold $c$ is set to $0.85$.
\end{itemize}

\subsection{The Simulated Sim2Snr1 Dataset from the QSM Challenge 2.0}
The QSM reconstruction challenge 2.0 (Seoul 2019) provides the simulated Sim2Snr1 dataset with a ground-truth susceptibility map for evaluation \cite{QSMChallenge:2019}. The dataset contains an intra-hemispheric calcification that creates strong dephasing effect and is thus more challenging. The simulation parameters are as follows: the main magnetic field $B_0=7$ T, repetition time (TR) = 50 ms; echo time (TE) TE1/TE2/TE3/TE4 = 4/12/20/28 ms; echo spacing = 8 ms; the flip angle = 15\textdegree; field of view (FoV) = 164$\times$205$\times$205 mm$^3$ and $1$ mm$^3$ isotropic voxels. The Gaussian noise is added to the complex data, producing a peak SNR of 100. Since only the brain tissues are used to simulate the field perturbations, the background field removal is not needed. 

As recommended in \cite{Milovic:L1QSM:2022,Milovic:PT_QSM:2021}, the L1-QSM and FANSI approaches used the provided frequency map with a simulated TE of 10 ms to recover the susceptibility map. As recommended by the MEDI-toolbox \cite{NonlinearMEDI:Liu:2013}, the MEDI approach used the provided frequency map with a simulated TE equalling the echo spacing (i.e. 8 ms) to recover the susceptibility map. In contrast, the AMP-PE approach used all the unwrapped phase images obtained from 4 echoes to recover the susceptibility map.

\subsection{\emph{in vivo} 3D Brain Dataset}
We acquired \emph{in vivo} 3D brain data from healthy subjects on a 3T MRI scanner (Prisma model, Siemens Healthcare, Erlangen, Germany), with the written consent obtained before imaging. In addition, we conducted a retrospective chart review study to obtain clinical MR images from patients with hemorrhage under the approval of the Institutional Review Board of Emory University. The patient scans were performed using a 3T MRI scanner (Tim Trio model, Siemens Healthcare, Erlangen, Germany). We reconstructed susceptibility maps from both the healthy subject scans and patient scans with brain hemorrhage. The data were acquired using GRE sequences with the following acquisition parameters:
\begin{itemize}
\item \emph{Five healthy subject scans (H1--H5)}. We have the main magnetic field $B_0=3$ T, the flip angle = $15$\textdegree, the number of echoes = 4, the first echo time = 7.32 ms, echo spacing = 8.68 ms, TR = 38 ms, slice thickness = 0.7 mm, in-plane resolution = 0.6875 mm $\times$ 0.6875 mm, bandwidth per pixel = 260 Hz, and acquisition matrix size = 320 $\times$ 320 $\times$ 208. 

\item \emph{One patient scan with hemorrhage (P1).} We have the main magnetic field $B_0=3$ T, the flip angle = $15$\textdegree, the number of echoes = 6, the first echo time = 7.5 ms, echo spacing = 7.6 ms, slice thickness = 1.8887 mm, in-plane resolution = 0.9375 mm $\times$ 0.9375 mm, and acquisition matrix size = 184 $\times$ 256 $\times$ 80.

\item \emph{Two patient scans with hemorrhage (P2,P3).} We have the main magnetic field $B_0=3$ T, the flip angle = $15$\textdegree, the number of echoes = 4, the first echo time = 6.35 ms, echo spacing = 6.05 ms, TR = 35 ms, slice thickness = 2 mm, in-plane resolution = 0.71875 mm $\times$ 0.71875 mm, and acquisition matrix size = 260 $\times$ 320 $\times$ 72.
\end{itemize}

The initial phase $\boldsymbol\phi_0$ was estimated using the complex fitting method available in the MEDI toolbox, and then removed from the raw phase images. The resulting multi-echo phase images were unwrapped using the 3D best-path phase unwrapping algorithm \cite{Abdul-Rahman:3dbestpath:2007}. After unwrapping, the multi-echo phase images were divided by their respective echo times and combined to generate an average total field (in Hz). The averaged total field was then transformed to the phase domain (in radians) using a TE that equals the echo-spacing, resulting in a combined phase image. To remove the background field from the combined phase image, we utilized the projection onto dipole fields (PDF) method \cite{Liu:PDF:2011}. The processed phase image (after background field removal) was the converted to the corresponding local field map (in Hz). In order to attain optimal performance from the L1-QSM, FANSI, and MEDI approaches, the local field map was mapped to the phase image (in radians) at a simulated TE to recover the susceptibility map. As recommended by the FANSI toolbox, a TE of $20$ ms was used for the \emph{in vivo} reconstruction by the L1-QSM and FANSI approaches. The MEDI approach, on the other hand, used the echo spacing as the recommended TE. However, the echo spacing may vary among different acquisition protocols. In order to verify that MEDI achieved optimal performance under the current experimental setting, we conducted experiments where the simulated TE varied from $1$ ms to $20$ ms. 

On the other hand, as shown in Fig. \ref{fig:factor_graph_qsm}, AMP-PE naturally supports the use of multi-echo phase images as measurements. For the AMP-PE approach, instead of processing a combined phase image, we applied PDF on the unwrapped multi-echo phase images individually to remove the background field. We then used the processed multi-echo phase images from all the echoes directly to recover the susceptibility as before. 

During the background field removal and the linear/nonlinear dipole inversion, a phase-based quality mask was also applied to remove voxels with unreliable phase values \cite{Karsa:2020}. In particular, if there were holes inside the brain mask, we need to fill them during background field removal, and reintroduce the holes during dipole inversion.


\subsection{An investigation of the optimal settings of AMP-PE}
\label{subsec:methods_settings_amp_pe}
As previously discussed, the optimal settings for AMP-PE differ between the simulated Sim2Snr1 dataset and the \emph{in vivo} datasets. Specifically, we investigated the following three settings of AMP-PE:
\begin{enumerate}[label=\arabic*)]
\item The choice of whether to enforce the ROI mask on $\boldsymbol\chi$ by setting the susceptibility outside the ROI to $0$ during dipole inversion.
\item The choice of the sparsifying wavelet basis.
\item The choice of the percentage threshold $c$ of the morphology mask $\mathcal{M}_v$ for wavelet coefficients.
\end{enumerate}
We showcase the effect of each choice through the following three experiments:
\begin{enumerate}[label=\arabic*)]
\item We compare the recovered susceptibility maps with and without the ROI mask enforced on $\boldsymbol\chi$. For the Sim2Snr1 dataset, we used the db1 wavelet basis, and set the percentage threshold $c$ to $75\%$. For the \emph{in vivo} dataset H1, we used the db2 wavelet basis, and set the percentage threshold to $85\%$.
\item We compare the recovered susceptibility maps using different wavelet bases. For the Sim2Snr1 dataset, we enforced the ROI mask on $\boldsymbol\chi$, set the percentage threshold $c$ to $75\%$, and selected wavelet bases from db1 and db2. For the \emph{in vivo} dataset H1, we did not enforce the ROI mask on $\boldsymbol\chi$, set the percentage threshold $c$ to $85\%$, and selected wavelet bases from db1, db2, db3, and db6.
\item We compare the recovered susceptibility maps using different percentage thresholds $c$. For the Sim2Snr1 dataset, we enforced the ROI mask on $\boldsymbol\chi$, used the db1 wavelet basis, and selected percentage thresholds from $50\%$, $75\%$, $80\%$, and $85\%$. For the \emph{in vivo} dataset H1, we did not enforce the ROI mask on $\boldsymbol\chi$, used the db2 wavelet basis, and selected percentage thresholds from $80\%$, $85\%$, $90\%$, and $95\%$.
\end{enumerate}

\section{Results}
\label{sec:results}
\subsection{The Simulated Sim2Snr1 Dataset from the QSM challenge 2.0}

\begin{table}[tp]
\caption{The simulated Sim2Snr1 dataset: evaluation metric scores of the recovered susceptibility maps using different approaches. The best performance under each metric is in boldface.}
\vspace{0.5em}
\label{tab:Sim2Snr1_evaluation_metric}
\centering
\resizebox{\columnwidth}{!}{
\begin{tabular}{llllllllll}
\toprule
 & & \multicolumn{3}{c}{\bfseries detrend NRMSE} & & & & \\ \cmidrule(lr){3-5}  
\multirow{-2}{*}{\bfseries Methods} &\multirow{-2}{*}{\bfseries NRMSE} & Tissue & Blood &DGM &\multirow{-2}{*}{\bfseries CalcStreak} &\multirow{-2}{*}{\bfseries DFCM} &\multirow{-2}{*}{\bfseries SSIM} &\multirow{-2}{*}{\bfseries HFEN} \\ \midrule
L1-QSM (L-curve) &51.01 &50.86 &150.29 &29.74 &8.17e-2 &28.21 &0.550 &58.85 \\
L1-QSM (Visual) &35.01 &33.23 &91.50 &18.08 &3.21e-2 &15.03 &0.795 &33.58\\
FANSI (L-curve) & 30.93 & \bf{32.99} & 65.43 & \bf{17.90} & \bf{1.00e-2} & 12.79  &0.790 & 29.67\\
MEDI (Default) & 35.13 & 33.35 & 81.75 & 20.60 & 1.95e-2 & 11.21 & 0.775 & \bf{25.28} \\
AMP-PE & \bf{30.86} & 33.48 & \bf{64.95} & 18.15 & 1.30e-2 & \bf{7.74} & \bf{0.811} & 28.91 \\
\bottomrule
\end{tabular}
}
\end{table}

\begin{figure}[tbp]
\centering
\includegraphics[width=\textwidth]{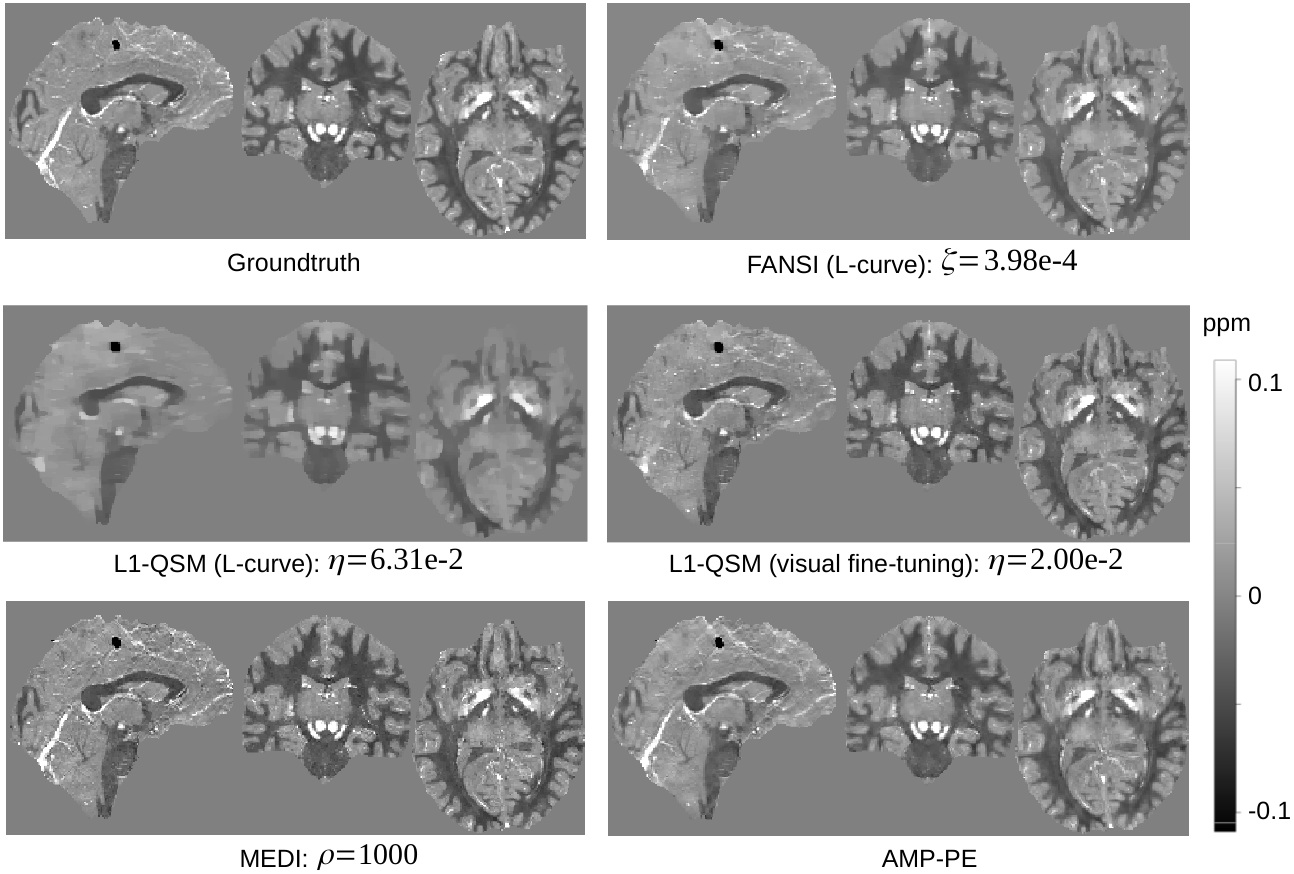}
\caption{The Sim2Snr1 dataset: recovered susceptibility maps using the L1-QSM, nonlinear FANSI, MEDI, and AMP-PE approaches.} 
\label{fig:Sim2Snr1_qsm}
\end{figure}

Using the code provided by the QSM challenge 2.0, we computed several evaluation metrics, including the normalized root mean square error (NRMSE), ROI-based detrend error metrics \footnote{The detrend NRMSE compensates for potential ``systematic underestimation'' and ``demeaning global shifts'' in susceptibility within the region considered.} for tissue, blood, and deep gray matter (DGM), calcification error metrics that include calcification streaking (CalcStreak) and deviation from calcification moment (DFCM), structural similarity index measure (SSIM), and high-frequency error norm (HFEN). The evaluation metric scores are shown in Table \ref{tab:Sim2Snr1_evaluation_metric}, and the reconstructed QSMs are shown in Fig. \ref{fig:Sim2Snr1_qsm}. The corresponding error maps are shown in Fig. S2 of the Supporting Information. In general, the FANSI and AMP-PE approaches outperform the L1-QSM approach. In terms of global metrics, AMP-PE achieves the lowest NRMSE and the highest SSIM, MEDI achieves the lowest HFEN. In terms of region-specific metrics, each approach has its own strong suit: FANSI achieves the lowest detrend NRMSE on the tissue and DGM, and the lowest CalcStreak; AMP-PE achieves the lowest detrend NRMSE on the blood, and the lowest DFCM.


Following the guidelines in \cite{Milovic:PT_QSM:2021}, we used the heuristic L-curve method to select the parameters for the L1-QSM and FANSI approaches respectively. The corresponding curvature curves are shown in Fig. S21 and S30 in the Supporting Information. As shown in Fig. \ref{fig:Sim2Snr1_qsm}, the parameter $\eta$ of L1-QSM selected by the L-curve method was not optimal and over-regularized the recovered susceptibility map, visual fine-tuning was further employed to find a working parameter for L1-QSM.

\subsection{\emph{in vivo} 3D Brain Dataset}
\begin{figure}[tbp]
\centering
\includegraphics[width=\textwidth]{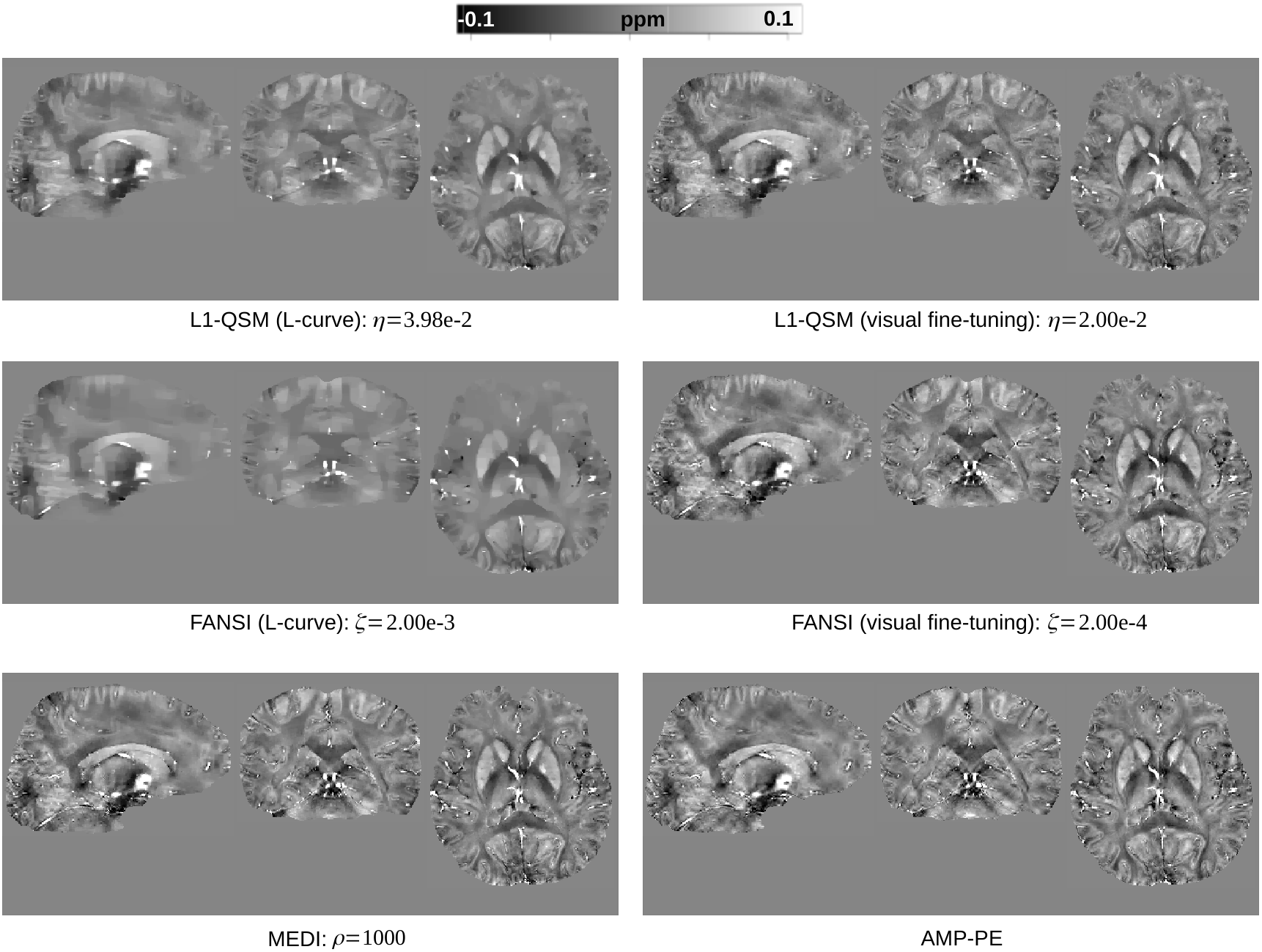}
\caption{Healthy scan (H1): recovered susceptibility maps using the L1-QSM approach, nonlinear FANSI, MEDI, and AMP-PE with the db2 wavelet basis.} 
\label{fig:invivo_healthy_scan_H1}
\end{figure}

\begin{figure}[tbp]
\centering
\includegraphics[width=\textwidth]{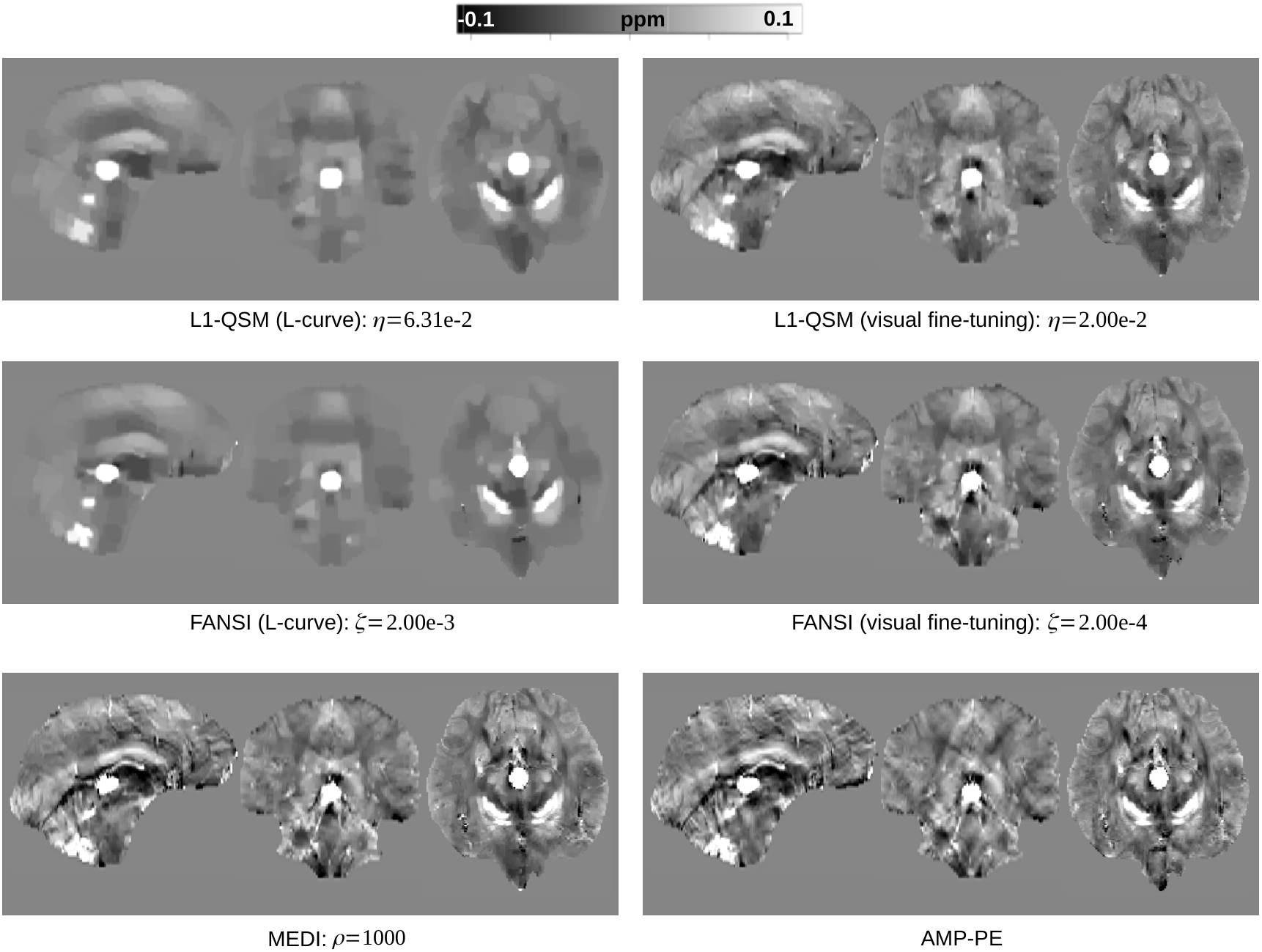}
\caption{Patient scan (P1): recovered susceptibility maps using the L1-QSM approach, nonlinear FANSI, MEDI, and AMP-PE with the db2 wavelet basis.} 
\label{fig:invivo_patient_scan_P1}
\end{figure}

\begin{figure}[tbp]
\centering
\includegraphics[width=\textwidth]{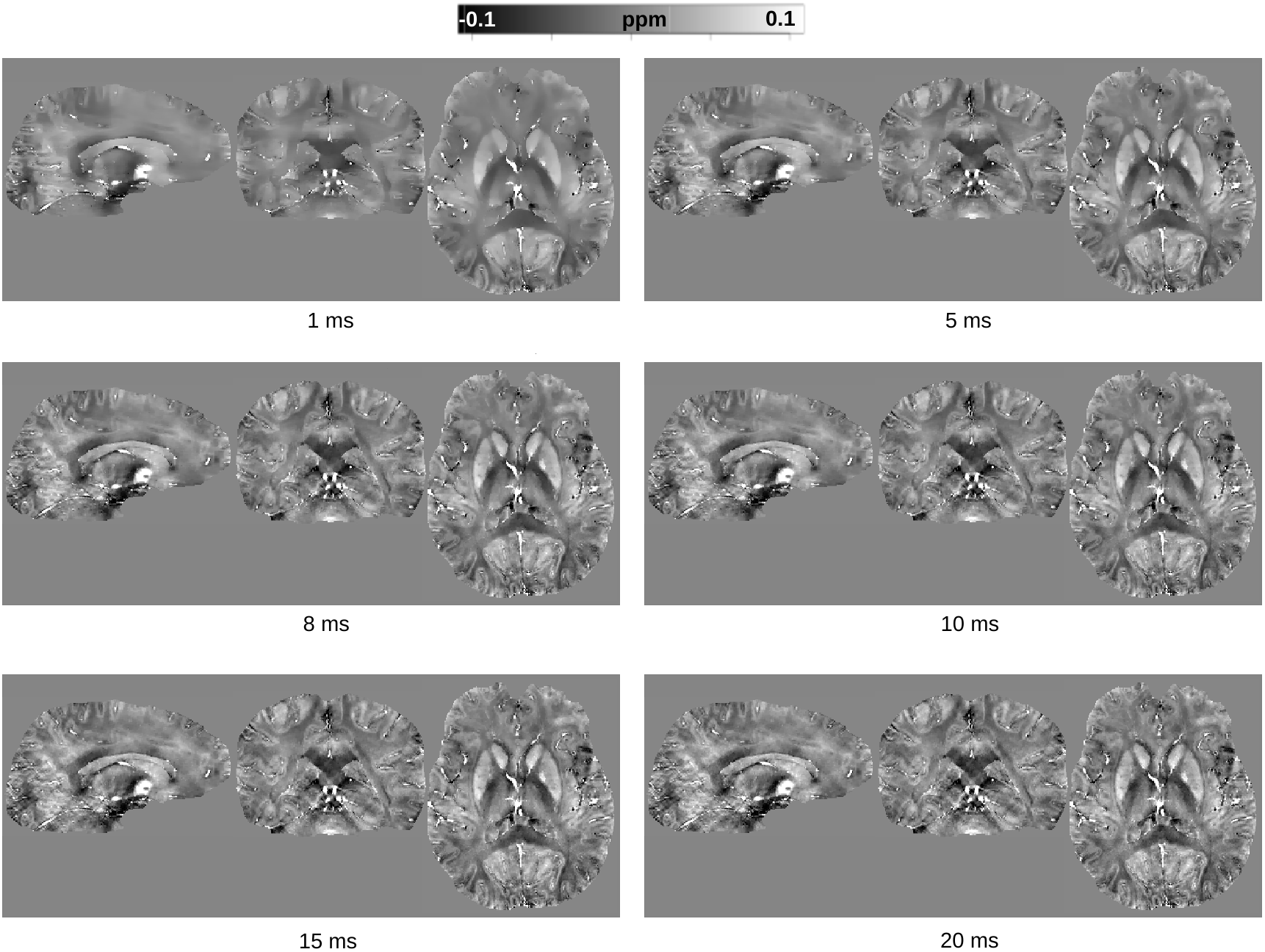}
\caption{The recovered susceptibility maps using the MEDI approach when the local field (in Hz) is mapped to the phase images at different echo times. Through visual fine-tuning, we can see that the default parameter $\rho=1000$ produces the best results when the chosen TE is set to $8\sim 10$ ms.} 
\label{fig:invivo_healthy_medi_echo}
\end{figure}

The recovered susceptibility maps for the healthy scan ``H1'' and hemorrhage scan ``P1'' are shown in Fig. \ref{fig:invivo_healthy_scan_H1}--\ref{fig:invivo_patient_scan_P1}. Additional QSMs from ``H2--H5'' and ``P2-P3'' are show in Fig. S3-S6 and S7-S8 in the Supporting Information. For the L1-QSM and FANSI approaches, although the L-curve method proved effective on the simulated Sim2Snr1 dataset, the parameters derived from this method over-regularized the reconstructions on the \emph{in vivo} datasets, leading to a loss of finer details in the recovered maps. The corresponding curvature curves are shown in Fig. S22-S29 and S31-S38 in the Supporting Information. As suggested in \cite{Milovic:FANSI:2018}, visual fine-tuning was employed to determine working parameters when the L-curve method failed. However, the parameters determined by visual fine-tuning are subjective and dependent on the practitioner. For the MEDI approach, the default parameter $\rho=1000$ was used. Fig. \ref{fig:invivo_healthy_medi_echo} shows the susceptibility maps recovered by MEDI when the local field (in Hz) is mapped to the phase images at different echo times. To ensure the best performance of MEDI, we need to double-check the results through visual fine-tuning as well. We can see that the default parameter produces the best results when the chosen TE is set to $8\sim 10$ms, which is consistent with the current experimental setting. On the other hand, the proposed AMP-PE recovers the susceptibility maps directly from the processed multi-echo phase images after phase unwrapping and background field removal. AMP-PE estimates the parameters from the data automatically and adaptively, with no need for visual fine-tuning.

The reconstructions of susceptibility maps are performed on the MATLAB platform using a machine (Intel Xeon Gold 5218 Processor, 2.30GHz) with 200 Gb RAM. We compared the runtime of each method on the healthy scan ``H1''. The L1-QSM and FANSI approaches took 4.7 and 3.5 minutes respectively for a single run, they took a total of 94.3 and 85.8 minutes respectively to complete the l-curve analysis of 20 and 25 parameter choices. The MEDI and AMP-PE approaches took 11.2 and 53.2 minutes respectively to complete the reconstruction.

\subsection{An investigation of the optimal settings of AMP-PE}
For the Sim2Snr1 dataset, the recovered susceptibility maps using AMP-PE under the three settings in Section \ref{subsec:methods_settings_amp_pe} are shown in Fig. S26-S27 in the Supporting Information, the corresponding evaluation metric scores are given in Table S2 in the Supporting Information. We can see that the best results are obtained when the ROI mask is enforced on $\boldsymbol\chi$, the db1 wavelet basis is used, and the percentage threshold $c$ is set to $75\%$.

\begin{figure}[tbp]
\centering
\label{fig:tr002_compare_roi_mask}
\subfigure[The choice of enforcing the ROI mask on the susceptibility map]{
\includegraphics[width=\textwidth]{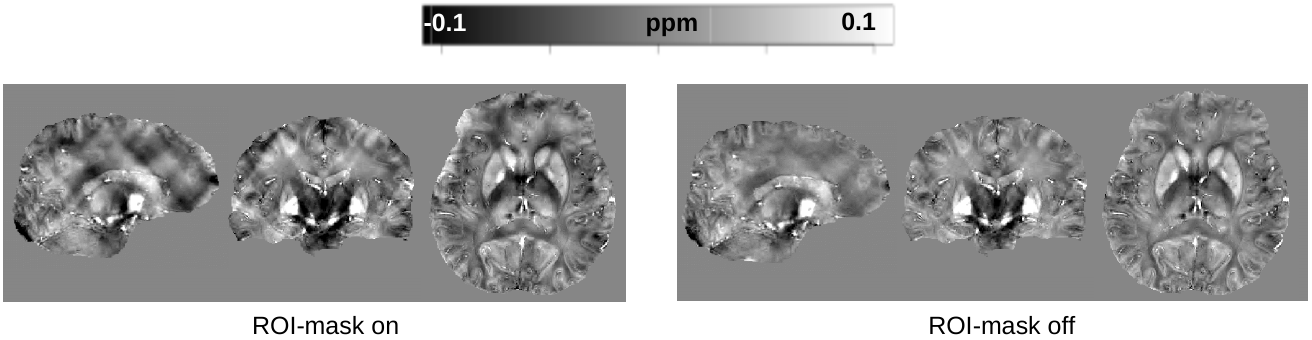}}\\
\label{fig:tr002_compare_wavelet_bases}
\subfigure[The choice of different wavelet bases]{
\includegraphics[width=\textwidth]{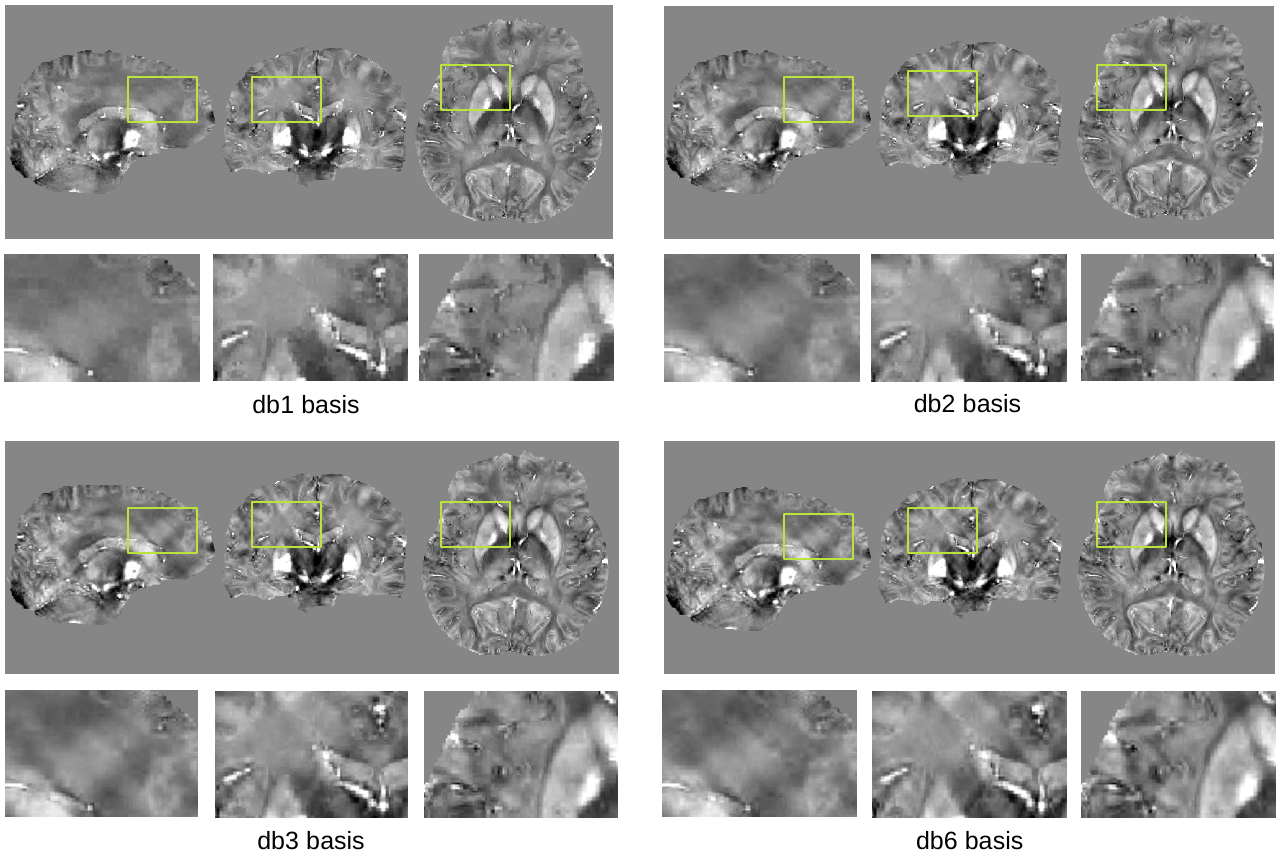}}\\
\caption{The H1 dataset: the recovered susceptibility maps and their error maps using the AMP-PE approach with different choices of settings.} 
\label{fig:tr002_qsm_compare_roi_mask_wavelt_bases}
\end{figure}

\begin{figure}[tbp]
\centering
\subfigure{
\includegraphics[width=\textwidth]{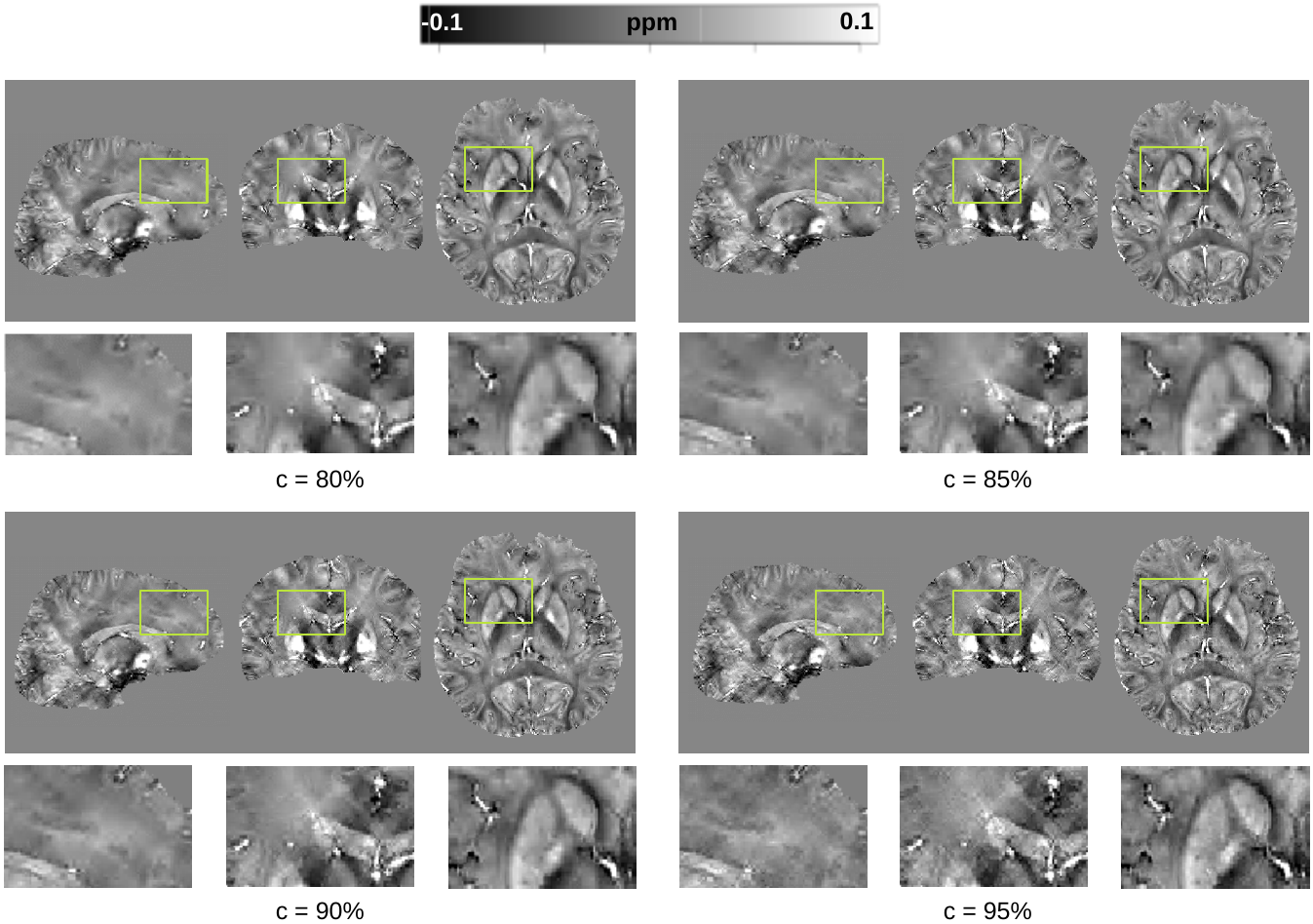}}
\caption{The H1 dataset: the recovered susceptibility maps and their error maps using the AMP-PE approach when the percentage threshold $c$ of the morphology mask $\mathcal{M}_g$ is chosen from $80\%$, $85\%$, $90\%$ and $95\%$.} 
\label{fig:tr002_qsm_compare_percentage_threshold}
\end{figure}

For the \emph{in vivo} dataset H1, the recovered susceptibility maps using AMP-PE under the three settings in Section \ref{subsec:methods_settings_amp_pe} are shown in Fig. \ref{fig:tr002_qsm_compare_roi_mask_wavelt_bases}-\ref{fig:tr002_qsm_compare_percentage_threshold} respectively. Due to the absence of ground-truth in the \emph{in vivo} case, we assessed the recovered susceptibility maps through visual inspection. We can see that the best results are obtained when the ROI mask is ``not'' enforced on $\boldsymbol\chi$, the db2 wavelet basis is used, and the percentage threshold $c$ is set to $85\%$.

\section{Discussion}

The L-curve method has previously been used to compute the regularization parameter. However, due to its heuristic nature, it is not robust enough to handle varying signal and noise conditions. The L-curve was initially proposed for the least-squares data-fidelity term with Tikhonov regularization (i.e. the squared $l_2$-norm) and achieved relatively robust performance \cite{Hansen:l_curve:2000}. However, when it was used for the $l_1$-norm regularization in FANSI, the computed parameter may not be close to the optimal parameter, since the statistical property of the $l_1$-norm is different from that of the squared $l_2$-norm. To address this issue, the L-curve method was combined with visual fine-tuning in \cite{Milovic:L1QSM:2022} for \emph{in vivo} reconstructions. This requires human intervention and produces subjective results. In this paper, we propose the Bayesian approach for QSM reconstruction that allows us to estimate the parameters adaptively and automatically. The estimated parameters maximize their posterior distributions, which avoids the subjectivity introduced by visual fine-tuning.

For the simulated Sim2Snr1 dataset, the local field map is derived solely from the brain tissue within the ROI. As shown in Fig. S39(a) and Table S2, it is therefore beneficial to enforce the binary brain ROI-mask on $\boldsymbol\chi$ during the dipole inversion. However, for practical \emph{in vivo} reconstructions, the background field removal process would not be perfect, and there is still some residue background field left within the ROI. As shown in Fig. \ref{fig:tr002_compare_roi_mask}, enforcing the ROI-mask on $\boldsymbol\chi$ leads to incorrect susceptibility variations. By allowing the susceptibility outside the ROI to account for the residual background field, we can obtain a significantly improved susceptibility map in the \emph{in vivo} case.

The forward model under the AMP framework is essentially constructed with respect to the transform coefficients $\vv$ of the susceptiblity map $\boldsymbol\chi$. To recover $\boldsymbol\chi$ from $\vv$, the sparsifying transform applied to the image must be invertible. In this paper, we employ the wavelet transform to obtain the sparse representation of the image in the form of wavelet coefficients. Although the total-variation transform is not invertible and therefore cannot be utilized, the db1 wavelet shrinkage with a single level is equivalent to a single step of total-variation regularization \cite{Steidl:2005}. Fig. \ref{fig:tr002_qsm_compare_roi_mask_wavelt_bases} shows the susceptibility maps recovered using AMP-PE with various wavelet bases. As the wavelet basis order increases, more high-frequency information can be captured. The db1 basis cannot capture enough high-frequency information, leading to pixelation effects in the recovered map. In contrast, the db6 basis captures high-frequency information from both the image and streaking artifacts (see the coronal views in Fig. \ref{fig:tr002_qsm_compare_roi_mask_wavelt_bases}). For the \emph{in vivo} reconstruction, we can either simply employ the db2 basis for reconstruction or average the susceptibility maps obtained using the db1 and db2 bases if resources permit. However, we must note that the db2 basis may not be the optimal choice for other MRI tasks. For example, in the case of $T_1$ and $T_2^*$ mappings, which do not involve the removal of streaking artifacts, the db6 basis was used in \cite{Bayesian_R2Star_2022} to obtain the optimal reconstruction of $T_2^*$ maps.

The use of a morphology mask $\mathcal{M}_g$ in TV regularization was first proposed by Liu et al. in the MEDI approach \cite{Liu:MEDI:2012}, it can also be used in other TV regularization approaches, such as FANSI. By penalizing non-edge gradients, the structural information can be incorporated through the morphology mask $\mathcal{M}_g$. The edges in the susceptibility map recovered by MEDI are thus better preserved, which leads to a significantly lower HFEN. Inspired by this, we proposed a morphology mask $\mathcal{M}_v$ that can be applied on the wavelet coefficients in Section \ref{subsec:mask_wavelet_coeff}. This mask enables us to separate the high-frequency information of structural edges from that of streaking artifacts in the wavelet domain. As shown in Fig. \ref{fig:tr002_qsm_compare_percentage_threshold}, we can set the percentage threshold $c$ at around $85\%$ to generate a suitable mask for the \emph{in vivo} reconstruction.

As shown in Fig. \ref{fig:gm_noise}, we employ the two-component Gaussian-mixture distribution to model the noise in QSM. The second component of the mixture model is used to handle noise outliers caused by phase unwrapping errors or brain hemorrhage and calcification. In the Supporting Information, Fig. S41(a) shows a comparison of the recovered susceptibility maps when we used the (one-component) Gaussian distribution and the (two-component) Gaussian-mixture to model the noise in the case of phase unwrapping errors. We can see that AMP-PE with a Gaussian-mixture noise prior is better at removing the streaking artifact than AMP-PE with a single-Gaussian noise prior. As discussed in Section \ref{subsec:amp_pe}, the ill-posed dipole kernel causes AMP-PE to overestimate the weight $\xi_2$ of the second Gaussian component. To address this problem, we propose a two-step procedure whereby $\xi_1,\xi_2$ are first estimated based on a preliminary reconstruction and then fixed in the final reconstruction. The left plot of Fig. S41(b) shows the recovered map by the standard procedure where $\xi_1,\xi_2$ are freely updated; and the right plot of Fig. S41(b) shows the recovered map by the proposed two-step procedure where the streaking artifacts are removed more effectively and the images are cleaner.

The convergence of AMP has been established for random Gaussian measurement matrices. However, for non-Gaussian measurement matrices, damping and mean-removal operations are necessary to ensure convergence. Since the means of rows of the transform operator $\mA_e$ in \eqref{eq:linear_measurement} are approximately zero, the mean of the resulting measurement operator in \eqref{eq:exp_measurement_model_linearized} is also approximately zero. Therefore, we do not need to perform the mean-removal operation in this case. However, the distribution $p(\mA_e)$ of the entries in $\mA_e$ is not Gaussian. For example, in the simple case where the size of $\mA_e$ is $10^3\times 10^3$, the normal probability plot in Fig. S42 of the Supporting Information shows that $p(\mA_e)$ is both left- and right-skewed with discontinuity in the distribution, and the histogram in Fig. S42 confirms that $p(\mA_e)$ is quite different from the Gaussian distribution. As discussed in Section \ref{subsec:amp_pe}, we used damping operations on the estimated susceptibility and parameters to achieve the convergence of AMP. Furthermore, since the QSM reconstruction is a nonconvex problem, initialization also plays a critical role in stabilizing the AMP algorithm. Our experiments have shown that good performance can be achieved when the susceptibility was initialized with a zero vector, while the parameters were initialized by performing a simple maximum-likelihood fitting of the least-squares solution.

\section{Conclusion}
We propose a probabilistic Bayesian formulation to recover susceptibility maps from nonlinear complex exponential measurements that are robust to phase errors. To model the noise outliers, we adopt a custom two-component Gaussian mixture noise prior. Our approach also employs a sparsity-promoting Laplace signal prior on the image wavelet coefficients to improve the image quality. We use the proposed AMP with automatic and adaptive parameter estimation (AMP-PE) to recover the susceptibility map. Additionally, we introduce a morphology mask on the image wavelet coefficients to incorporate anatomical structural information into the reconstruction. Our \emph{in vivo} experiments demonstrate that AMP-PE is robust and successfully recovers susceptibility maps with estimated parameters. Whereas the L1-QSM, FANSI and MEDI methods typically rely on visual fine-tuning to select or double-check working parameters for different MR protocols and scanners. The proposed AMP-PE is equipped with built-in parameter estimation and avoids the subjectivity from the visual fine-tuning step, making it an excellent choice for the clinical setting.




{\bfseries \Large DATA AVAILABILITY STATEMENT}

The source code from this paper is openly available at: \urlstyle{tt}\url{https://github.com/EmoryCN2L/QSM_AMP_PE}

\bibliography{ref}

\newpage

{\bfseries\Large Supporting Information}

Additional Supporting Information may be found online in the Supporting Information section.

\paragraph{Supporting Table S1} The simulated Sim2Snr1 dataset: evaluation metric scores of the recovered susceptibility maps using the linear recovery approach.

\paragraph{Supporting Table S2} The simulated Sim2Snr1 dataset: evaluation metric scores of the recovered susceptibility maps using AMP-PE under different settings.

\paragraph{Supporting Figure S1} The factor graph of the sparse signal recovery task under the AMP framework: ``$\bigcirc$'' represents the variable node, and ``$\blacksquare$'' represents the factor node.

\paragraph{Supporting Figure S2} The Sim2Snr1 dataset: the error maps using the L1-QSM, FANSI, MEDI, and AMP-PE approaches.

\paragraph{Supporting Figures S3--S6} Healthy scan (H2--H5): recovered susceptibility maps using the L1-QSM approach, nonlinear FANSI, MEDI, and AMP-PE with the db2 wavelet basis.

\paragraph{Supporting Figures S7,S8} Patient scan (P2, P3): recovered susceptibility maps using the L1-QSM approach, nonlinear FANSI, MEDI, and AMP-PE with the db2 wavelet basis.

\paragraph{Supporting Figure S9} The Sim2Snr1 dataset: recovered susceptibility map and its error maps using the linear recovery approach.

\paragraph{Supporting Figure S10, S11} The recovered susceptibility maps using the linear recovery approach with different choices of settings.

\paragraph{Supporting Figures S12--S20} The Sim2Snr1 and \emph{in vivo} datastes: the linear recovery approach. Left subfigure: the L-curve that shows the log-fidelity-cost vs. the log-regularization-cost. Right subfigure: the curvatures on the L-curve with respect to the corresponding regularization weights. The median-filtered curvatures are used to determine the inflection point where the curvature changes its sign.

\paragraph{Supporting Figures S21--S29} The Sim2Snr1 and \emph{in vivo} datastes:  The L1-QSM approach. Left subfigure: the L-curve that shows the log-fidelity-cost vs. the log-regularization-cost. Right subfigure: the curvatures on the L-curve with respect to the corresponding regularization weights. The median-filtered curvatures
are used to determine the inflection point where the curvature changes its sign.

\paragraph{Supporting Figures S30--S38} The Sim2Snr1 and \emph{in vivo} datastes: the nonlinear FANSI approach. Left subfigure: the L-curve that shows the log-fidelity-cost vs. the log-regularization-cost. Right subfigure: the curvatures on the L-curve with respect to the corresponding regularization weights. The median-filtered curvatures are used to determine the inflection point where the curvature changes its sign.

\paragraph{Supporting Figure S39} The Sim2Snr1 dataset: the recovered susceptibility maps and their error maps using the AMP-PE approach with different choices of settings: (a) the choice of enforcing the ROI mask on the susceptibility map; (b) the choice of different wavelet bases.

\paragraph{Supporting Figure S40} The Sim2Snr1 dataset: the recovered susceptibility maps and their error maps using the AMP-PE approach when the percentage threshold $c$ of the morphology mask $\mathcal{M}_g$ is chosen from $50\%$, $75\%$, $80\%$ and $85\%$.

\paragraph{Supporting Figure S41} (a) The Gaussian-mixture noise prior is able to model the noise outliers and thus better at removing the streaking artifacts from recovered susceptibility map; (b) AMP-PE achieves better performance with the two-step procedure where the Gaussian-mixture weights are first estimated based on a preliminary reconstruction and then fixed during then final reconstruction.

\paragraph{Supporting Figure S42} The distribution of the entries in the transform operator $\mA_e$: (a) The normal probability plot compares the sample distribution to the normal distribution: all the $N$ sample entries are sorted and plotted on the $x$-axis, and the $y$-axis represents the corresponding quantiles of the distribution with respect to each entry. For the $i$-th sorted entry on the $x$-axis, its $y$-axis value is $\frac{i-0.5}{N}$; (b) The histogram of the $N$ entries.

\end{document}